\documentclass[aps,prd,twocolumn,superscriptaddress,nofootinbib]{revtex4-2}
\usepackage{amsmath,amssymb,bm}
\usepackage{graphicx}
\usepackage{booktabs}
\usepackage{siunitx}
\usepackage{xcolor}

\usepackage{multirow}
\graphicspath{{figures/}}

\newcommand{\dd}{\mathrm d}
\newcommand{\GeV}{\mathrm{GeV}}
\newcommand{\TeV}{\mathrm{TeV}}
\newcommand{\keV}{\mathrm{keV}}

\begin{document}

\title{\textcolor{black}{A Dark-Dimension Origin of Geometric Inelastic Dark Matter: The LUX-ZEPLIN High-Recoil Event and Multi-Target Tests}}

\author{Waqas Ahmed}
\affiliation{Center for Fundamental Physics, School of Artificial Intelligence, Hubei Polytechnic University, Huangshi 435003, China}
\email{waqasmit@hbpu.edu.cn}
\author{George K. Leontaris}
\email{leonta@uoi.gr}
\affiliation{
Physics Department, Theory Division,
University of Ioannina,
45110 Ioannina, Greece}

\begin{abstract}
The extended nuclear-recoil analysis of LUX-ZEPLIN (LZ) has reported one event compatible with a recoil energy near $248$ keV, motivating endothermic dark-matter scenarios with support at large recoil energy. We study a pseudo-Dirac electroweak-doublet dark-matter model embedded in a five-dimensional Dark-Dimension framework. The Standard Model and the vectorlike electroweak doublets are localized on the visible brane, while dark-number violation is communicated through a neutral bulk state. The resulting mass splitting is exponentially suppressed,
$\delta=\delta_{\rm UV}e^{-2\pi MR}$,
providing a geometric origin for the few-hundred-keV scale relevant to inelastic scattering. The neutral weak current is dominantly off diagonal, fixing the neutron-scale normalization at
$\sigma_{\chi n}^{Z}\simeq1.86\times10^{-39}\,\mathrm{cm^2}$.
Using a two-bin Poisson diagnostic of the extended LZ recoil window, we find a shallow minimum near
$(m_\chi,\delta)\simeq(1.35~\TeV,366~\keV)$, while the thermally motivated point
$(1.1~\TeV,360~\keV)$ lies only $\Delta\chi^2\simeq0.21$ higher. The same benchmark is kinematically inaccessible to light targets but remains open for xenon and tungsten, providing a characteristic multi-target test of the model. For CaWO$_4$ we obtain an integrated tungsten rate of
$3.15\times10^{-2}$ events kg$^{-1}$ yr$^{-1}$ in the $95$--$150$ keV interval. These results connect the geometric origin of the pseudo-Dirac splitting with high-recoil direct-detection phenomenology.

\end{abstract}

\maketitle

\section{Introduction}

The simplest weakly interacting dark-matter searches are built around elastic scattering and a recoil spectrum concentrated at low energy.  Endothermic dark matter is different.  If the stable halo state $\chi_1$ can scatter only by converting into a slightly heavier state $\chi_2$, the inelastic threshold removes the low-velocity part of the halo and moves the observable signal toward larger recoil energy \cite{TuckerSmith2001,TuckerSmith2005,Bramante2016}.  This makes the high-recoil tail an unusually sensitive probe of a small mass splitting.

The recent LZ extended-energy analysis has made that observation timely.  With an exposure of 2.84 tonne-years and a nuclear-recoil window extending to approximately 270 keV, LZ reported one event consistent with a nuclear recoil of $248\pm23_{\rm stat}\pm23_{\rm sys}$ keV.  The maximum local significance is $3.4\sigma$ and the global significance, after the look-elsewhere effect, is $2.6\sigma$ \cite{LZ2026}.  The event is therefore not evidence for dark matter.  It is, however, located in precisely the part of phase space where conventional elastic WIMPs are sparse and endothermic scattering can become distinctive.

Several interpretations appeared immediately after the LZ result.  Su, Yang and Yang showed that splittings of order a few hundred keV and dark-matter masses above several hundred GeV can place spectral support in the new LZ window \cite{Su2026}.  Higgsino interpretations are particularly economical because the neutral Higgsino states already form a pseudo-Dirac system and their $Z$ interaction is dominantly off diagonal \cite{Freese2026,FanReece2026,WuZhangZhu2026,Yin2026}.  In that case the scattering strength is fixed by the weak interaction rather than tuned to reproduce a single event.

The question addressed here is where such a small splitting could come from.  The required scale, $\delta\sim\mathcal O(10^{-4})$ GeV, is tiny compared with a TeV electroweak-doublet mass, but it need not be a fundamental low-energy input.  Extra-dimensional ``geography'' can communicate symmetry breaking between separated branes with an exponentially small amplitude \cite{ArkaniHamedDimopoulos2002,ArkaniHamedDvaliMarchRussell2001}.  We use that idea in a form adapted to the Dark Dimension.

The Dark-Dimension proposal relates the small observed vacuum-energy density to a mesoscopic compact dimension \cite{MonteroVafaValenzuela2023}.  Writing the vacuum-energy density as $\rho_\Lambda$, its characteristic size is
\begin{equation}
R=\lambda\rho_\Lambda^{-1/4},
\end{equation}
so the first Kaluza--Klein scale is $m_{\rm KK}=R^{-1}=\lambda^{-1}\rho_\Lambda^{1/4}$.  A micron-size benchmark corresponds to a sub-eV KK spacing and a higher-dimensional species scale around $10^9$--$10^{10}$ GeV \cite{MonteroVafaValenzuela2023}.  Recent Dark-Dimension constructions explicitly place the Standard Model on a codimension-one brane while allowing neutral fermions -in the keV range-  and gauge sectors to propagate in the bulk \cite{MonteroVafaValenzuela2026}. 

This observation also tells us how the present dark-matter model should be embedded.  If the electroweak doublets themselves propagated in a micron-size dimension, their charged and neutral KK excitations would be separated only by the sub-eV/eV compactification scale.  Here instead, we  place the Standard Model and the TeV vectorlike doublets on the visible brane and let only an electroweak-singlet messenger probe the Dark Dimension, whilst the dark-number violation is localized at the distant brane.  The messenger propagates across the interval and back, so the induced Majorana mass of the brane dark matter is suppressed by the square of a brane-to-brane propagator.  At low momentum this gives
\begin{equation}
\delta\simeq\delta_{\rm UV}e^{-2\pi M R},
\end{equation}
where $M$ is the neutral bulk mass and $\delta_{\rm UV}$ collects the brane couplings, the distant-brane symmetry-breaking insertion and higher-dimensional normalization factors.  In this form the observed dark-energy scale fixes the geometric scale $R$, while direct detection constrains the dimensionless separation parameter $M R$.


A second purpose of this paper is to move beyond an LZ-only interpretation.  A few-hundred-keV endothermic threshold creates a strong hierarchy among detector targets.  Light nuclei simply cannot supply the required reduced mass, while iodine, xenon and especially tungsten remain accessible.  This observation connects the LZ event to existing inelastic searches with PICO \cite{PICO2023} and to the planned high-energy reach of CRESST \cite{CRESSTUpgrade,Su2026}.  It also explains why conventional XENONnT and PandaX WIMP analyses, despite large exposures, do not automatically test the same region: the benchmark xenon spectrum begins near 200 keV, substantially above the recoil region for which standard WIMP searches are optimized \cite{XENONnT2025,PandaX2021}.

The paper is organized as follows.  Section II develops the Dark-Dimension embedding and derives the pseudo-Dirac splitting.  Section III formulates the weak inelastic recoil rate.  Section IV introduces the LZ likelihood diagnostic and presents the xenon benchmark.  Section V develops the multi-target kinematics and the xenon--tungsten comparison.  Section VI discusses astrophysical, nuclear, detector and ultraviolet-model systematics.  Section VII summarizes our conclusions.
\section{Dark-Dimension scale and geometric origin of the splitting}
\label{sec:model}

The Dark-Dimension proposal identifies the small observed vacuum energy with the decompactification scale of one mesoscopic extra dimension \cite{MonteroVafaValenzuela2023}.  We denote the four-dimensional vacuum-energy density by $\rho_\Lambda$ and keep the dimensionless proportionality factor explicit,
\begin{equation}
R=\lambda\rho_\Lambda^{-1/4}.
\label{eq:Rdark}
\end{equation}
The corresponding KK spacing is
\begin{equation}
m_{\rm KK}=R^{-1}
=\lambda^{-1}\rho_\Lambda^{1/4}.
\label{eq:mKK}
\end{equation}
Using $\rho_\Lambda^{1/4}\simeq2.25$ meV, a convenient micron benchmark is
\begin{align}
R&\simeq0.88~\mu{\rm m}
\left(\frac{\lambda}{10^{-2}}\right)
\left(\frac{2.25~{\rm meV}}{\rho_\Lambda^{1/4}}\right),
\label{eq:Rnum}
\\
m_{\rm KK}&\simeq0.225~{\rm eV}
\left(\frac{10^{-2}}{\lambda}\right)
\left(\frac{\rho_\Lambda^{1/4}}{2.25~{\rm meV}}\right).
\label{eq:mkknum}
\end{align}
Up to order-one compactification conventions, the four- and five-dimensional Planck scales satisfy
\begin{equation}
M_{\rm Pl}^2\sim M_5^3R,
\end{equation}
so that
\begin{equation}
M_5\sim\left(M_{\rm Pl}^2m_{\rm KK}\right)^{1/3}.
\label{eq:M5}
\end{equation}
For the benchmark in Eq.~(\ref{eq:mkknum}) this is of order $10^9$ GeV, while the range commonly discussed in the Dark-Dimension literature extends to $10^{9}$--$10^{10}$ GeV \cite{MonteroVafaValenzuela2023}.

\begin{figure}[t]
\includegraphics[width=\columnwidth]{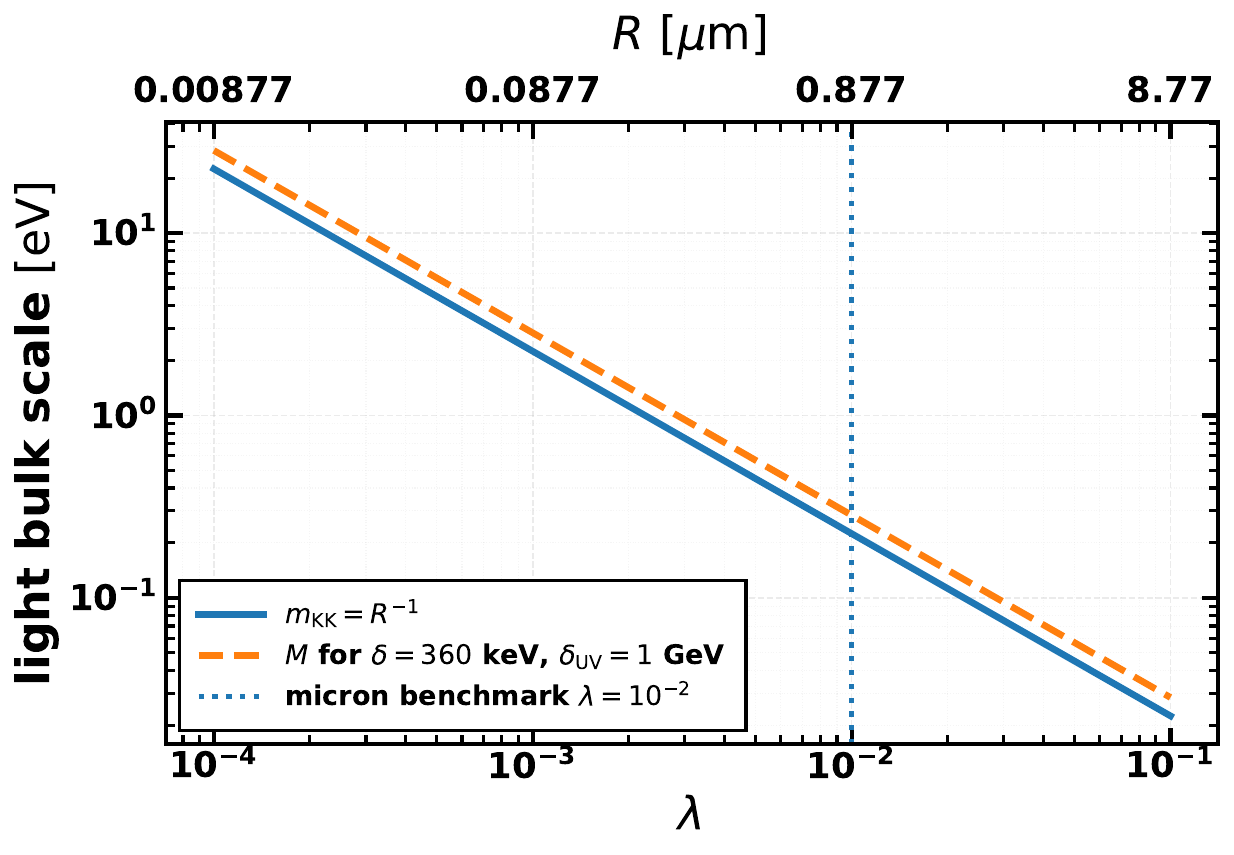}
\caption{KK scale implied by $R=\lambda\rho_\Lambda^{-1/4}$ for $\rho_\Lambda^{1/4}=2.25$ meV.  The dashed curve shows the neutral messenger mass corresponding to the illustrative $\delta=360$ keV, $\delta_{\rm UV}=1$ GeV benchmark.  The upper axis gives the associated compactification radius.}
\label{fig:ddscales}
\end{figure}

Figure~\ref{fig:ddscales} displays the relation between the dark-energy prefactor, the KK scale and the physical size of the extra dimension.  The relation is important conceptually: $\rho_\Lambda$ fixes the scale of the fifth dimension before the dark-matter model is introduced.

\subsection{Electroweak doublets are brane localized}

The low-energy dark sector contains two vectorlike electroweak doublets,
\begin{equation}
D_1\sim({\bf 2},-1/2),\qquad
D_2\sim({\bf 2},+1/2),
\label{eq:doublets}
\end{equation}
which are odd under a stabilizing dark parity.  Their neutral components are denoted by $N_1$ and $N_2$.  In the absence of dark-number violation the leading mass term is
\begin{equation}
{\cal L}_{0}\supset-\mu_D D_1\!\cdot\!D_2+{\rm H.c.},
\qquad \mu_D\sim{\cal O}({\rm TeV}).
\label{eq:DiracMass}
\end{equation}

\begin{figure}[t]
\includegraphics[width=\columnwidth]{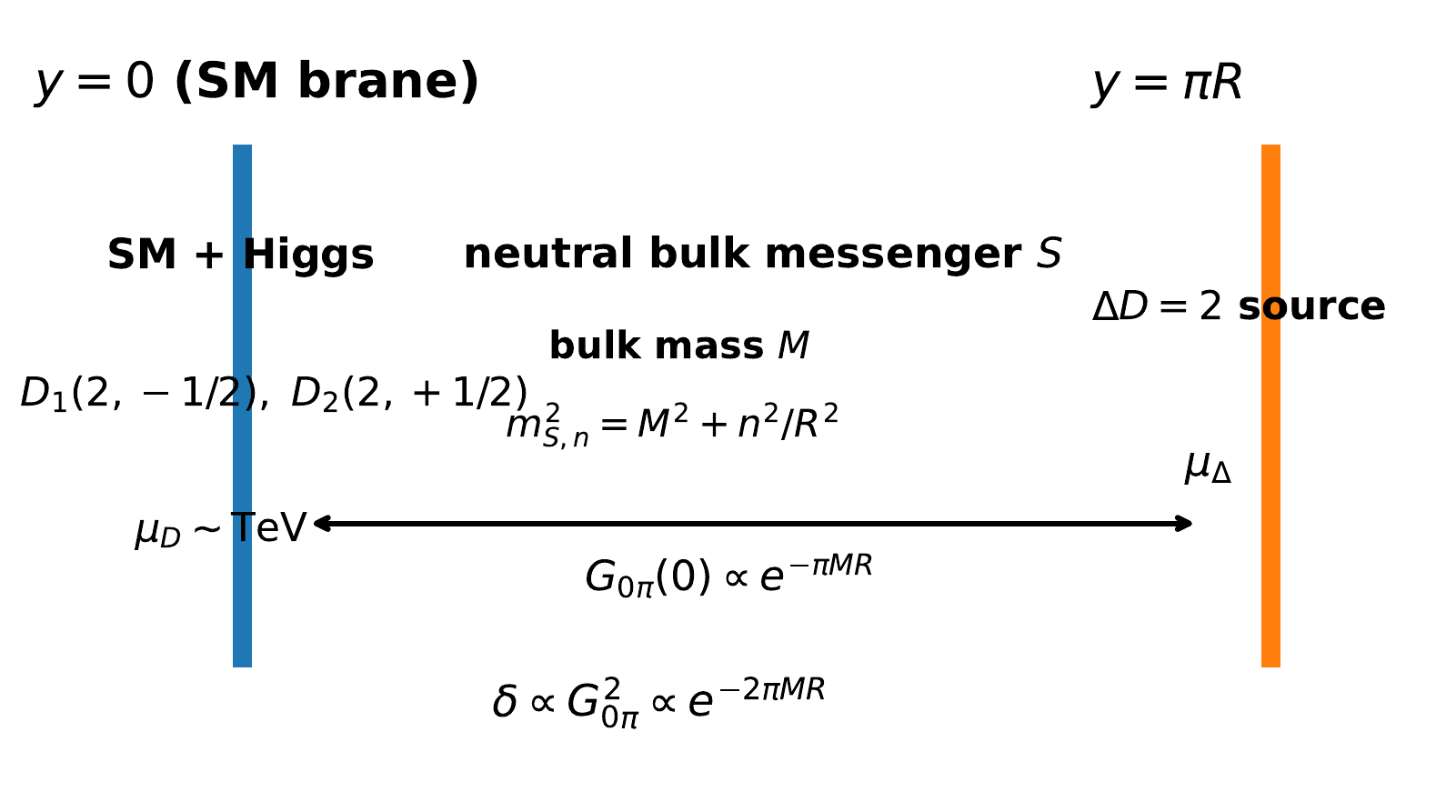}
\caption{Minimal Dark-Dimension embedding.  The Standard Model and the TeV electroweak doublets live at $y=0$.  A neutral bulk messenger communicates a dark-number-violating source from the distant brane.  Because the messenger must traverse the interval twice, the induced brane Majorana mass is proportional to $e^{-2\pi MR}$.}
\label{fig:setup}
\end{figure}

A direct identification of a bulk electroweak-doublet model with the micron Dark Dimension would be problematic.  The charged and neutral doublets would inherit KK excitations with spacing $m_{\rm KK}\ll{\rm GeV}$, producing a dense tower of electroweakly charged states.  We therefore adopt the brane--bulk structure shown in Fig.~\ref{fig:setup}: the Standard Model, the Higgs and $D_{1,2}$ are localized at $y=0$, whereas the fifth dimension is probed only by an electroweak-singlet messenger $S$.  This is in the same general spirit as recent Dark-Dimension constructions in which the Standard Model is brane localized while neutral fermions propagate in the bulk \cite{MonteroVafaValenzuela2026,AntoniadisChatrabhutiIsono2026}.

\subsection{Distant-brane dark-number violation}

Gauge invariance permits the visible-brane couplings
\begin{equation}
{\cal L}_{y=0}\supset
-\lambda_1 D_1\!\cdot\!H\,S
-\lambda_2 D_2\!\cdot\!\widetilde H\,S
+{\rm H.c.},
\label{eq:braneYuk}
\end{equation}
where $\widetilde H=i\sigma_2H^\ast$ and higher-dimensional normalization factors are left implicit.  Dark number is assumed to be an excellent symmetry on the visible brane.  It is broken at the distant boundary by a localized insertion schematically of the form
\begin{equation}
{\cal L}_{y=\pi R}
\supset-\frac12\mu_\Delta SS+{\rm H.c.}
\label{eq:farbrane}
\end{equation}
or by an equivalent scalar vacuum expectation value carrying two units of dark number.

The mechanism is the familiar distant-brane communication of an approximate symmetry \cite{ArkaniHamedDimopoulos2002,ArkaniHamedDvaliMarchRussell2001}.  For a massive bulk field, the zero-four-momentum propagator between opposite boundaries has the generic form
\begin{equation}
G_{0\pi}(0)\propto
\frac{1}{M\sinh(\pi MR)}
\longrightarrow
\frac{2}{M}e^{-\pi MR},
\label{eq:propagator}
\end{equation}
where the first proportionality is boundary-condition dependent but the exponential asymptotics is universal for $MR\gtrsim1$.  The Majorana self-energy of the brane state contains two such propagations, one from the visible brane to the dark-number-violating boundary and one back.  After electroweak symmetry breaking, the induced neutral mass matrix can therefore be written as
\begin{equation}
{\cal M}_0=
\begin{pmatrix}
 m_1 & \mu_D\\
 \mu_D & m_2
\end{pmatrix},
\qquad
m_{1,2}\ll\mu_D,
\label{eq:massmatrix}
\end{equation}
with
\begin{equation}
m_{1,2}\propto e^{-2\pi MR}.
\end{equation}
All brane couplings, the distant-brane insertion, cutoff factors and non-exponential propagator normalization are absorbed into an effective unsuppressed scale $\delta_{\rm UV}$.  The two neutral mass eigenstates are then Majorana fermions forming a pseudo-Dirac pair with
\begin{equation}
\delta\equiv m_{\chi_2}-m_{\chi_1}
\simeq\delta_{\rm UV}e^{-2\pi MR}.
\label{eq:delta}
\end{equation}
This parameterization is the quantity used in the direct-detection analysis; it does not require us to commit to a particular microscopic realization of $\mu_\Delta$.

\begin{figure}[t]
\includegraphics[width=\columnwidth]{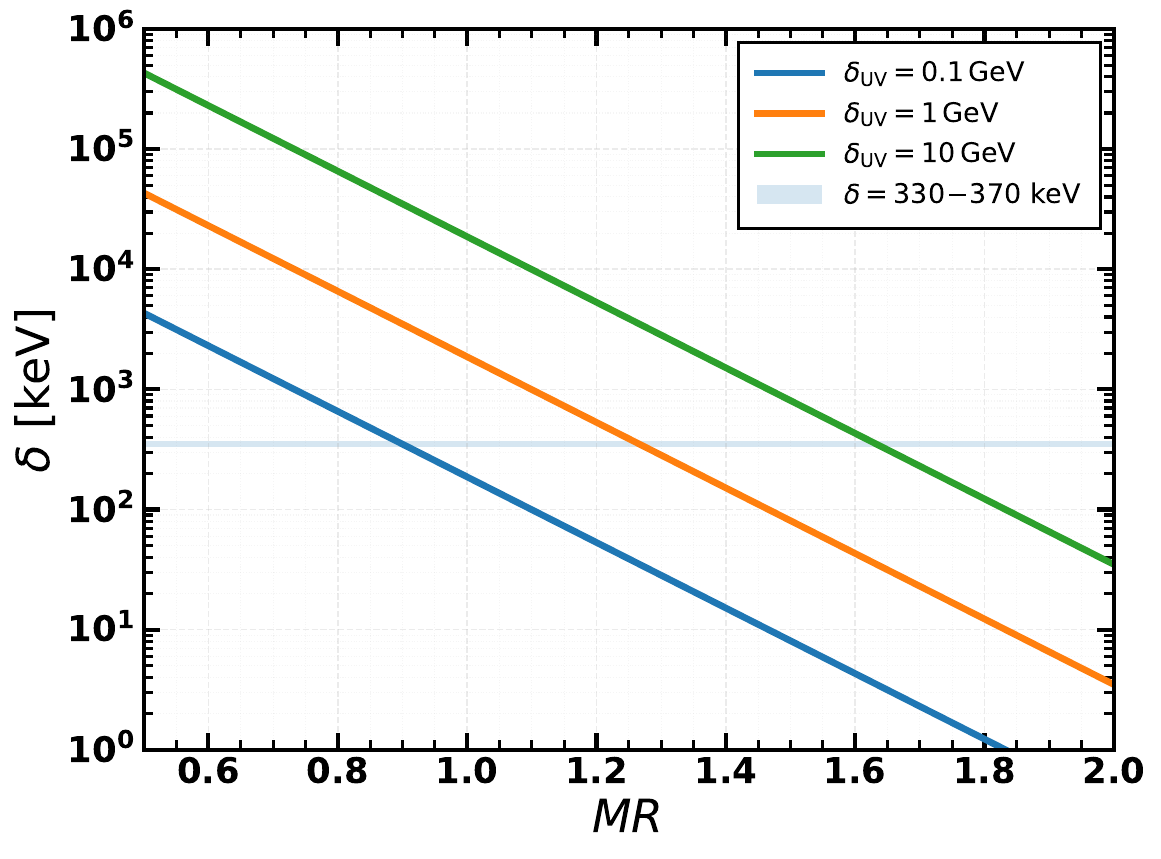}
\caption{Pseudo-Dirac splitting as a function of $MR=M/m_{\rm KK}$.  The shaded band highlights the few-hundred-keV region relevant to the LZ high-recoil event.  Once $R$ is fixed by the dark-energy scale, a direct-detection determination of $MR$ fixes the corresponding neutral bulk mass.}
\label{fig:geom}
\end{figure}

Equations~(\ref{eq:mKK}) and (\ref{eq:delta}) give the useful form
\begin{equation}
\delta=
\delta_{\rm UV}
\exp\!\left[-2\pi\frac{M}{m_{\rm KK}}\right].
\label{eq:deltaKK}
\end{equation}
Using the Dark-Dimension relation, this may also be written as
\begin{equation}
\delta=
\delta_{\rm UV}
\exp\!\left[-2\pi\lambda
\frac{M}{\rho_\Lambda^{1/4}}\right].
\label{eq:deltaLambda}
\end{equation}
Thus the observed vacuum-energy scale does not numerically predict $\delta$ by itself; it fixes $R$, while the LZ-sensitive splitting determines the dimensionless bulk parameter
\begin{equation}
MR=\frac{M}{m_{\rm KK}}
=\frac{1}{2\pi}\ln\!\left(\frac{\delta_{\rm UV}}{\delta}\right).
\label{eq:MRmap}
\end{equation}
For $\delta=360$ keV and $\delta_{\rm UV}=1$ GeV,
\begin{equation}
MR\simeq1.26.
\end{equation}
For the representative $\lambda=10^{-2}$ benchmark this corresponds to $m_{\rm KK}\simeq0.225$ eV and $M\simeq0.28$ eV.  The absolute messenger scale changes linearly with $m_{\rm KK}$, while the direct-detection prediction depends on the dimensionless combination $MR$.  Figure~\ref{fig:geom} shows the exponential map.

Because the KK tower is electroweak neutral, it does not face the immediate charged-state collider problem associated with bulk electroweak doublets.  Its mixing with the brane electroweak sector, thermal population and cosmological role do, however, depend on the brane coefficients in Eq.~(\ref{eq:braneYuk}) and constitute additional constraints on a complete ultraviolet model.  We therefore regard Eqs.~(\ref{eq:deltaKK})--(\ref{eq:deltaLambda}) as an effective Dark-Dimension embedding of the pseudo-Dirac splitting rather than as a fully specified cosmology of the singlet tower.


\subsection{Off-diagonal neutral current}

Before the Majorana perturbation is introduced, the neutral components form a Dirac fermion.  Rewriting the neutral current in the Majorana basis gives, to leading order in $m_{1,2}/\mu_D$,
\begin{equation}
{\cal L}_Z\supset
g_{Z12} Z_\mu\,
\overline{\chi_2}\gamma^\mu\chi_1+{\rm H.c.},
\label{eq:Zoff}
\end{equation}
where $g_{Z12}$ is fixed by the electroweak-doublet gauge quantum numbers; its phase and the explicit factor in a four-component Majorana convention are convention dependent. We therefore quote the standard zero-momentum inelastic-Higgsino normalization below, following Refs.~\cite{Bramante2016,Yin2026}.  The diagonal vector current vanishes at leading order.  Direct detection is therefore naturally inelastic,
\begin{equation}
\chi_1+N\rightarrow\chi_2+N.
\end{equation}

From this $Z$-mediated interaction, the resulting dark-matter--nucleon cross section is fixed by electroweak gauge couplings. 
For a neutron-scale normalization, we use


\begin{equation}
\sigma_{\chi n}^{Z}=\frac{G_F^2\mu_n^2}{8\pi}
\simeq1.86\times10^{-39}\,\mathrm{cm^2}
\label{eq:sigman}
\end{equation}
for TeV-scale dark matter.  This is the central predictive difference from generic inelastic models in which the nucleon cross section is scanned independently.

\section{Inelastic recoil rate}
\label{sec:rate}

For endothermic scattering the minimum speed required to produce a recoil $E_R$ is
\begin{equation}
v_{\min}(E_R)={\frac{1}{\sqrt{2m_AE_R}}\left(
\frac{m_AE_R}
{\mu_{\chi A}}+\delta\right)~,}
\label{eq:vmin}
\end{equation}
where $\mu_{\chi A}$ is the DM-nucleus reduced mass and $m_A$ the nucleus mass.
To find the minimum of this function, we impose
$\dd v_{\min}/\dd E_R=0$ and solve for $E_R$.
The minimum occurs at
\begin{equation}
E_R^\star=\delta\frac{\mu_{\chi A}}{m_A}.
\label{eq:estar}
\end{equation}

Equation~(\ref{eq:estar}) is useful when interpreting the location of a high-energy event.  For $m_\chi=1.1$ TeV and xenon, a recoil centered near 248 keV by itself corresponds to a splitting closer to $0.28$ MeV.  The larger $\sim0.35$--$0.37$ MeV values preferred by rate calculations arise because the normalization, the high-velocity tail, the nuclear response, and the finite analysis window all enter the likelihood.  This distinction is also emphasized in recent Higgsino analyses \cite{Yin2026}.
At the minimum the velocity is 
\begin{equation}
v_{\min}(E_R^\star)=\sqrt{\frac{2\delta}{\mu_{\chi A}}}
\end{equation}
This is the threshold velocity, i.e., the minimum speed required for endothermic scattering to occur at all. At this speed, the incoming DM particle has just enough kinetic energy to both excite the heavier state and produce the threshold recoil $E_R^\star$.

Subsequently, we use a truncated Standard Halo Model with
\begin{equation}
\rho_\chi=0.3~\GeV\,\mathrm{cm^{-3}},\quad
v_0=238~\mathrm{km\,s^{-1}},
\end{equation}
\begin{equation}
v_{\rm esc}=544~\mathrm{km\,s^{-1}},\quad
v_E=250.5~\mathrm{km\,s^{-1}}.
\end{equation}
The corresponding mean inverse speed is denoted by $\eta(v_{\min})$.

For completeness, the mean inverse speed is not a universal function: it depends on the assumed Galactic velocity distribution.  In this work we use the truncated Maxwell--Boltzmann Standard Halo Model, shifted into the Earth frame \cite{LewinSmith1996}.  In the Galactic frame,
\begin{equation}
\begin{aligned}
 f_{\rm G}(\mathbf u)&=
 \frac{e^{-u^2/v_0^2}}
 {N_{\rm esc}\,\pi^{3/2}v_0^3}
 \Theta(v_{\rm esc}-u),\\
 f_E(\mathbf v)&=f_{\rm G}(\mathbf v+\mathbf v_E).
\end{aligned}
\label{eq:velocitydist}
\end{equation}
with
\begin{equation}
 N_{\rm esc}=\operatorname{erf}(z)
 -\frac{2z}{\sqrt{\pi}}e^{-z^2},
 \qquad
 z\equiv\frac{v_{\rm esc}}{v_0}.
 \label{eq:Nesc}
\end{equation}
The quantity entering Eq.~(\ref{eq:rate}) is
\begin{equation}
 \eta(v_{\min})
 \equiv
 \int_{|\mathbf v|>v_{\min}}
 \frac{f_E(\mathbf v)}{v}\,d^3v.
 \label{eq:etadef}
\end{equation}
Defining
\begin{equation}
 x\equiv\frac{v_{\min}}{v_0},
 \qquad
 y\equiv\frac{v_E}{v_0},
 \qquad
 z\equiv\frac{v_{\rm esc}}{v_0},
\end{equation}
we use the standard analytic form
\begin{equation}
\eta(v_{\min})
=
\frac{1}{2N_{\rm esc}v_0y}\,
{\cal I}(x,y,z),
\label{eq:etaSHM}
\end{equation}
where
\begin{align}
{\cal I}(x,y,z)
={}&
\operatorname{erf}(x+y)
-\operatorname{erf}(x-y)
\nonumber\\
&-\frac{4y}{\sqrt{\pi}}e^{-z^2},
\qquad x<z-y ,
\label{eq:eta1}
\\[2mm]
{\cal I}(x,y,z)
={}&
\operatorname{erf}(z)
-\operatorname{erf}(x-y)
\nonumber\\
&-\frac{2(z-x+y)}{\sqrt{\pi}}e^{-z^2},
\nonumber\\[-1mm]
&\hspace{15mm} z-y\leq x<z+y ,
\label{eq:eta2}
\\[2mm]
{\cal I}(x,y,z)
={}&0,
\qquad x\geq z+y .
\label{eq:eta3}
\end{align}
Thus Eq.~(\ref{eq:etaSHM}) is the specific $\eta$ used in all recoil-rate calculations below.  A different halo model would imply a different mean inverse speed.  We keep $v_E=250.5~\mathrm{km\,s^{-1}}$ fixed in the baseline analysis and treat the resulting high-velocity-tail dependence as a systematic in Sec.~\ref{sec:systematics}.

For coherent weak scattering, with $s_W\equiv\sin\theta_W$ and $c_W\equiv\cos\theta_W$, the nuclear charge is
\begin{equation}
Q_W={\cal N}-(1-4s_W^2)Z,
\end{equation}
where, ${\cal N}=A-Z$ and the factor $(1-4s_W^2)Z$ comes from EW couplings of the $Z$-boson to protons and neutrons.
The zero-momentum cross section is
\begin{equation}
\sigma_A^0=
\frac{G_F^2\mu_{\chi A}^2}{8\pi}Q_W^2.
\label{eq:sigmaA}
\end{equation}
Since $1-4s_W^2\ll1$ at the relevant weak scale, the proton contribution is strongly suppressed relative to the neutron contribution, so that $Q_W\simeq{\cal N}$ to good accuracy.

The differential rate per unit target mass is then
\begin{equation} 
\frac{\dd R}{\dd E_R}
=N_T\frac{\rho_\chi}{m_\chi}
\frac{m_A\sigma_A^0}{2\mu_{\chi A}^2}
F_A^2(E_R)\,\eta(v_{\min}),
\label{eq:rate}
\end{equation}
where $N_T$ is the number of target nuclei per unit mass, $F_A$ is the nuclear form factor, and the appropriate factors of $c$ and unit conversion are included numerically.  Natural xenon is treated isotope by isotope using atom fractions.  We use a Helm form factor for the transparent baseline calculation.  At $E_R\sim250$ keV the momentum transfer is large, so the form-factor choice is one of the leading theory systematics; a final collaboration-level recast should use the one-body response employed by LZ.

\begin{figure}[t]
\includegraphics[width=\columnwidth]{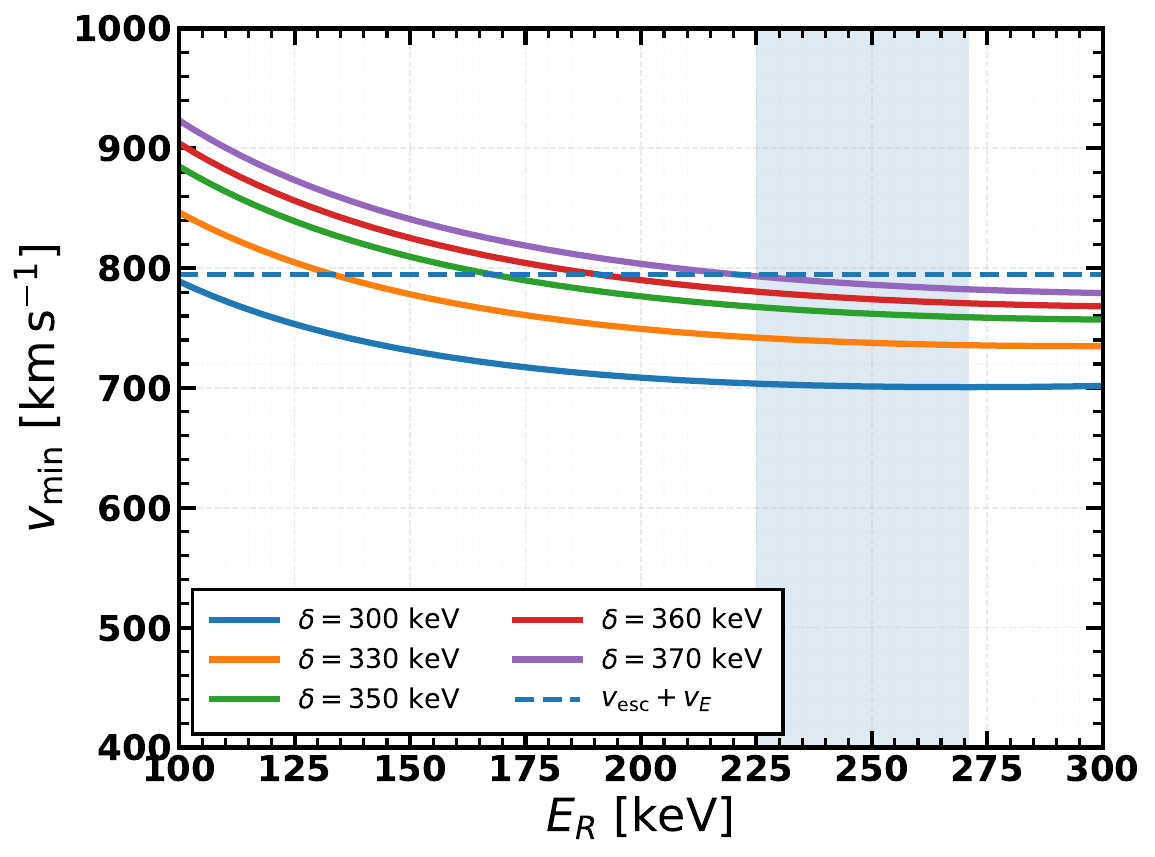}
\caption{Minimum incoming speed for $m_\chi=1.1$ TeV and representative splittings.  The shaded band denotes the $225$--$271$ keV high-energy interval used in the two-bin diagnostic.}
\label{fig:vmin}
\end{figure}

\begin{figure}[t]
\includegraphics[width=\columnwidth]{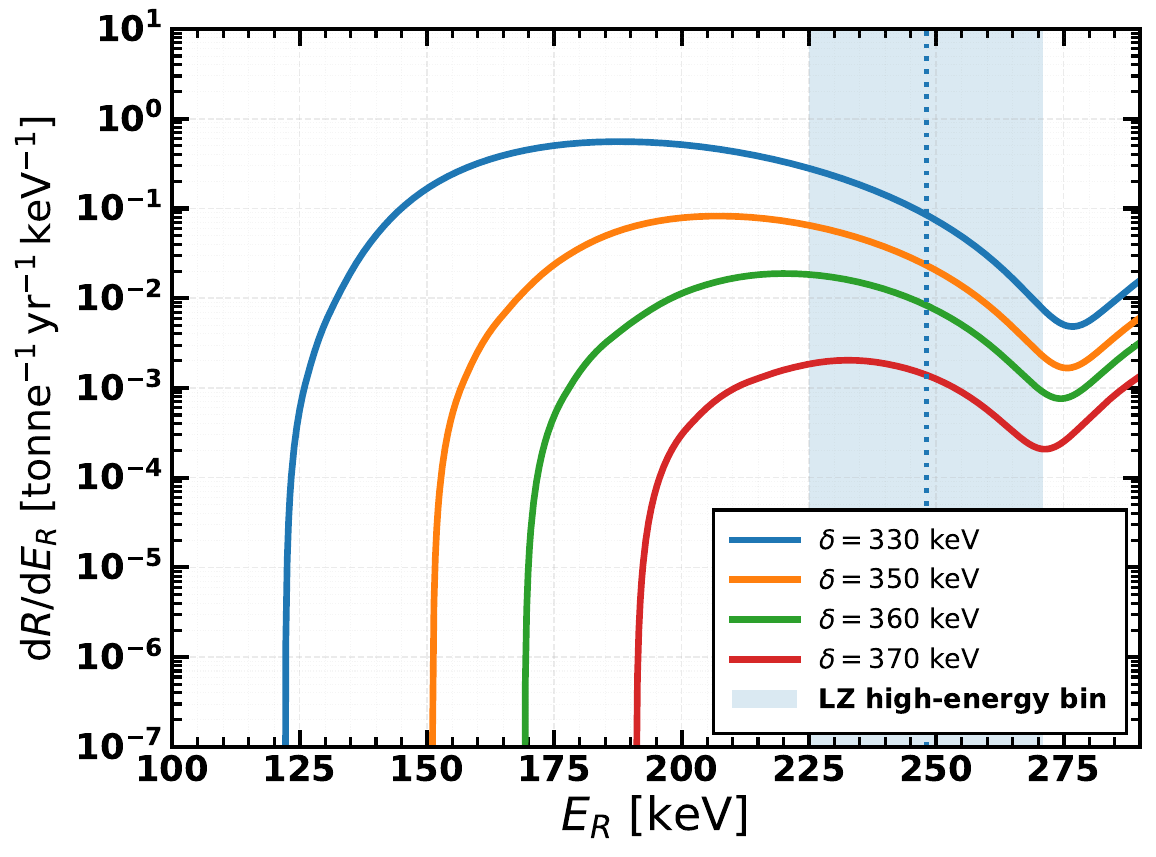}
\caption{Natural-xenon recoil spectra for the fixed weak interaction of Eq.~(\ref{eq:sigman}).  The inelastic threshold removes the low-energy signal as the splitting approaches the kinematic ceiling.}
\label{fig:xespec}
\end{figure}

Figure~\ref{fig:vmin} shows the rapid approach to the Galactic speed ceiling as $\delta$ increases.  Figure~\ref{fig:xespec} shows the corresponding xenon spectra.  Splittings near 360 keV strongly suppress low-energy events and concentrate the remaining rate near the high-energy edge of the LZ analysis.

\section{LZ statistical treatment}
\label{sec:lz}

The official LZ analysis is an event-level profile likelihood in detector observables with signal and background probability densities and nuisance parameters \cite{LZ2026}.  The public paper provides the likelihood structure, but our present calculation does not reproduce the complete detector-level workspace.  We therefore use a simpler likelihood only as a phenomenological diagnostic.

Following Ref.~\cite{Su2026}, we divide the high-recoil region into
\begin{equation}
160<E_R<200~\keV,\qquad n_1=0,
\end{equation}
and
\begin{equation}
225<E_R<271~\keV,\qquad n_2=1.
\end{equation}
For exposure ${\cal E}=2.84$ tonne yr, the expected count is
\begin{equation}
N_i={\cal E}\int_{E_i^{\rm min}}^{E_i^{\rm max}}
\dd E_R\,\epsilon(E_R)\frac{\dd R}{\dd E_R},\; i=1,2.
\label{eq:Ni}
\end{equation}
We used a smooth interpolation constrained by the published LZ efficiency endpoints for the executable baseline notebook.  Treating the two bins as independent Poisson counts and neglecting background uncertainties gives
\begin{equation}
{\cal L}=\prod_{i=1}^{2}\frac{N_i^{n_i}e^{-N_i}}{n_i!},
\qquad
\Delta\chi^2=-2\ln\frac{{\cal L}}{{\cal L}_{\rm max}}.
\label{eq:like}
\end{equation}
We stress that Eq.~(\ref{eq:like}) is the illustrative likelihood of Ref.~\cite{Su2026}, not the official LZ profile likelihood.

For our weak normalization, the numerical scan has a shallow minimum near
\begin{equation}
(m_\chi,\delta)_{\rm min}\simeq(1.35~\TeV,366~\keV),
\end{equation}
with
\begin{equation}
(N_1,N_2)\simeq(0.137,0.644).
\end{equation}
The thermally motivated electroweak-doublet point
\begin{equation}
(m_\chi,\delta)=(1.1~\TeV,360~\keV)
\label{eq:thermalbench}
\end{equation}
has
\begin{equation}
(N_1,N_2)\simeq(0.322,0.923),
\qquad
\Delta\chi^2\simeq0.21.
\end{equation}
With only one observed high-energy event, the difference is not statistically meaningful.  We therefore use Eq.~(\ref{eq:thermalbench}) as the principal benchmark for the multi-target study.

\begin{figure}[t]
\includegraphics[width=\columnwidth]{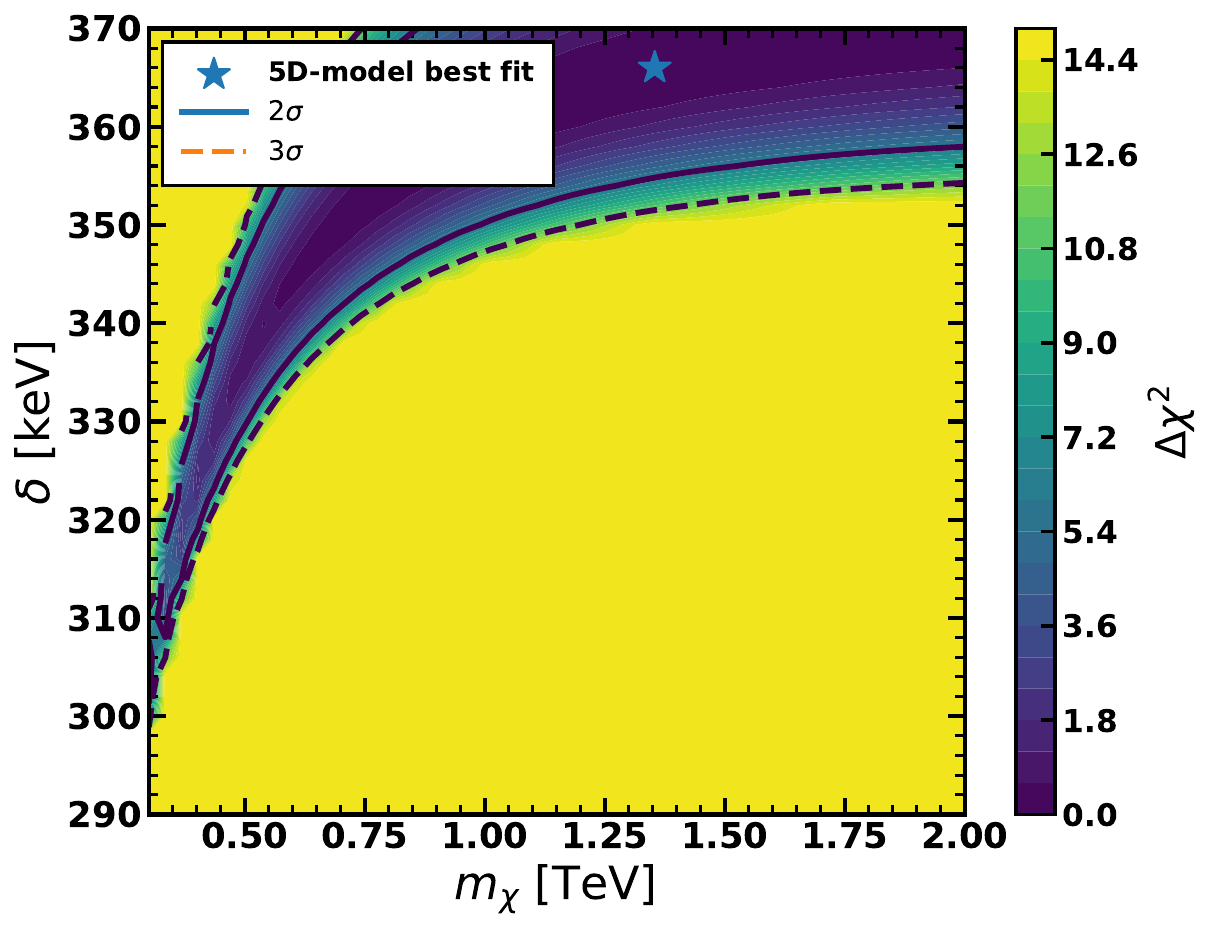}
\caption{Two-bin likelihood diagnostic in the $(m_\chi,\delta)$ plane for the fixed weak-interaction model.  The contours are based on Eq.~(\ref{eq:like}); they are not official LZ confidence regions.}
\label{fig:like}
\end{figure}

Figure~\ref{fig:like} summarizes the resulting shallow likelihood structure in the $(m_\chi,\delta)$ plane.  The broad preferred region reflects the limited information carried by a single high-energy event and should not be interpreted as an official LZ confidence region.

An independent normalization check is provided by Ref.~\cite{Freese2026}, which reports that for a TeV Higgsino and $\delta\simeq350$ keV the fixed weak vector cross section lies close to the two-sided LZ interval.  This is useful because the interaction strength in Eq.~(\ref{eq:sigman}) is not fitted to the event.

\section{Multi-target direct-detection tests}
\label{sec:targets}

For a target of mass $m_A$, endothermic scattering is possible only when
\begin{equation}
\delta\le\delta_{\rm max}(A)
=\frac12\mu_{\chi A}v_{\rm max}^2,
\label{eq:dmax}
\end{equation}
where we take $v_{\rm max}=v_{\rm esc}+v_E$.  This simple condition already determines which direct-detection experiments can test the benchmark.

\begin{figure}[t]
\includegraphics[width=\columnwidth]{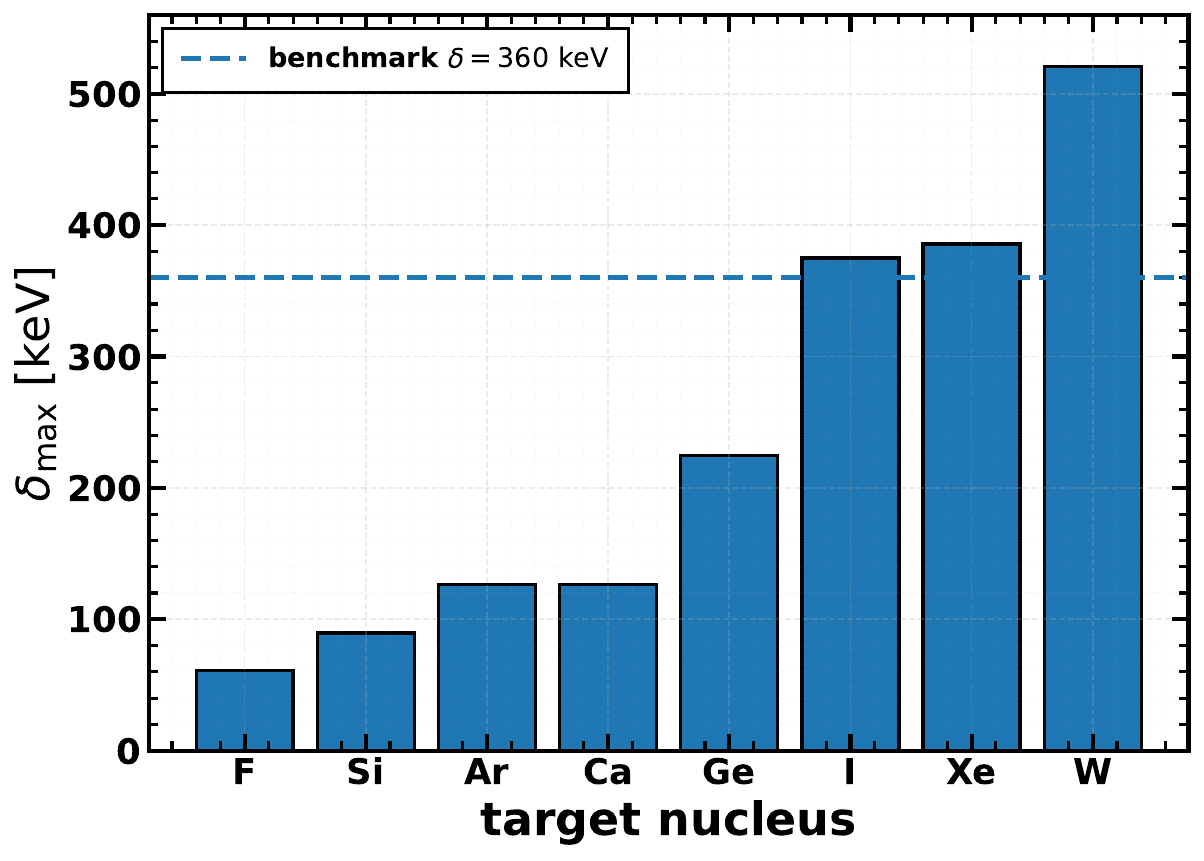}
\caption{Maximum endothermic splitting accessible to representative nuclei for $m_\chi=1.1$ TeV.  The horizontal line is the thermal benchmark $\delta=360$ keV.}
\label{fig:reach}
\end{figure}

For $m_\chi=1.1$ TeV, the maximal splittings are listed in Table~\ref{tab:reach} and shown in Fig.~\ref{fig:reach}.  A 360 keV transition is closed on F, Si, Ar, Ca and Ge, but remains open on I, Xe and W.  This immediately explains why light-target experiments are not competitive for the LZ-motivated splitting, regardless of exposure.

\begin{table}[t]
\caption{Kinematic reach at $m_\chi=1.1$ TeV for $v_{\rm max}=v_{\rm esc}+v_E=794.5$ km s$^{-1}$.}
\label{tab:reach}
\begin{ruledtabular}
\begin{tabular}{lccc}
Target & $A$ & $\delta_{\rm max}$ [keV] & $E_R$ window for $\delta=360$ keV [keV]\\
\hline
F  & 19  & 61  & closed\\
Si & 28  & 89  & closed\\
Ar/Ca & 40 & 127 & closed\\
Ge & 73 & 225 & closed\\
I  & 127 & 375 & 216--488\\
Xe & 131 & 386 & 191--550\\
W  & 184 & 521 & 89--1090\\
\end{tabular}
\end{ruledtabular}
\end{table}

\begin{figure}[t]
\includegraphics[width=\columnwidth]{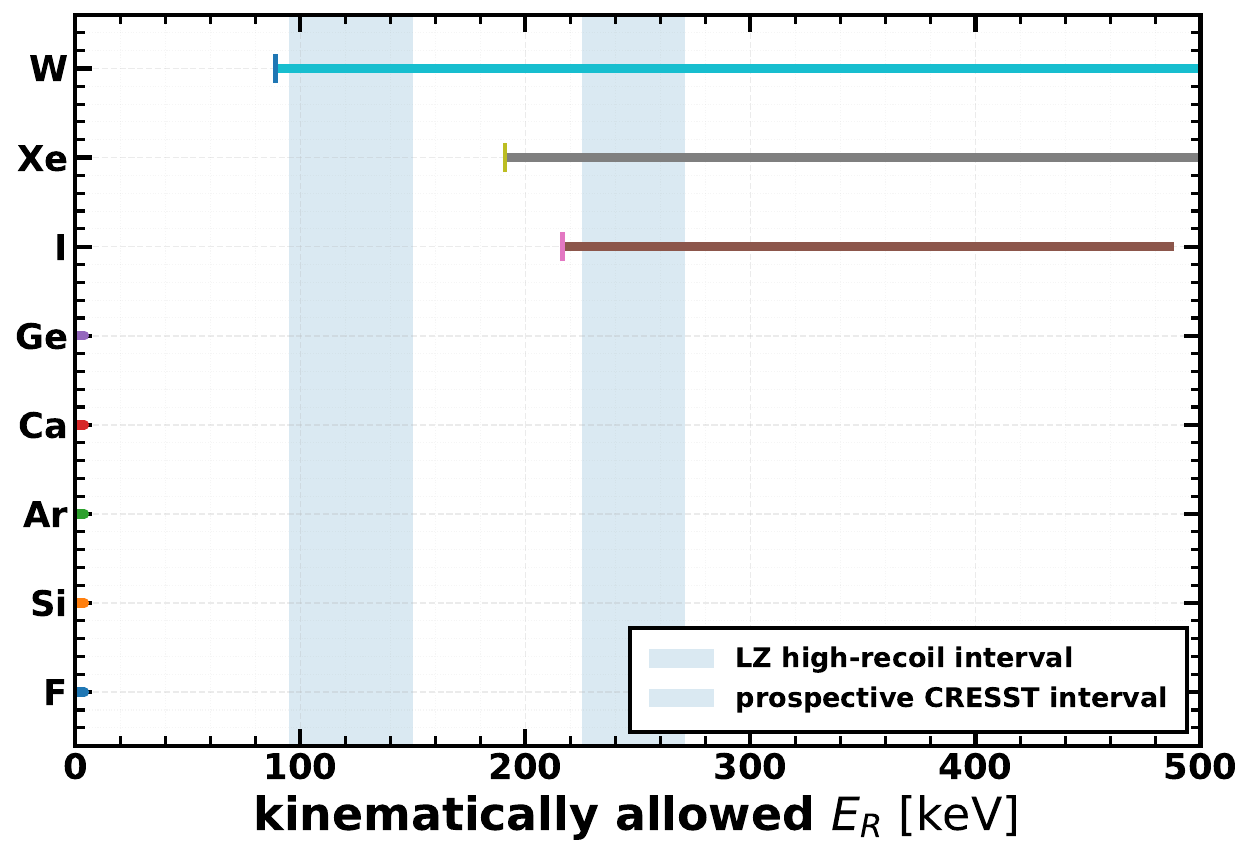}
\caption{Recoil intervals kinematically allowed at the benchmark $(m_\chi,\delta)=(1.1~\TeV,360~\keV)$.  Only I, Xe and W are open among the targets shown.  The LZ and prospective high-energy CRESST intervals are indicated for orientation.}
\label{fig:windows}
\end{figure}

The corresponding recoil intervals are displayed in Fig.~\ref{fig:windows}.  The figure makes the target hierarchy especially transparent: xenon opens only at high recoil energy, whereas tungsten retains a substantially lower recoil threshold.

Two consequences deserve emphasis.  First, the xenon lower endpoint is already $E_-\simeq191$ keV.  The standard XENONnT WIMP search, for example, is optimized from a few keV nuclear recoil upward and reports no excess in its conventional WIMP region \cite{XENONnT2025}; the older PandaX-4T commissioning analysis used an approximate 5--100 keV nuclear-recoil interval \cite{PandaX2021}.  Such searches do not automatically exclude the present benchmark because its signal is kinematically absent in their principal low-energy region.  An extended high-energy xenon analysis, of the type now performed by LZ, is essential.

Second, iodine is only marginally open, whereas tungsten has substantial phase space.  This is why PICO's dedicated inelastic program and a high-energy CRESST analysis are the most natural non-xenon tests \cite{PICO2023,Su2026}.

\begin{figure}[t]
\includegraphics[width=\columnwidth]{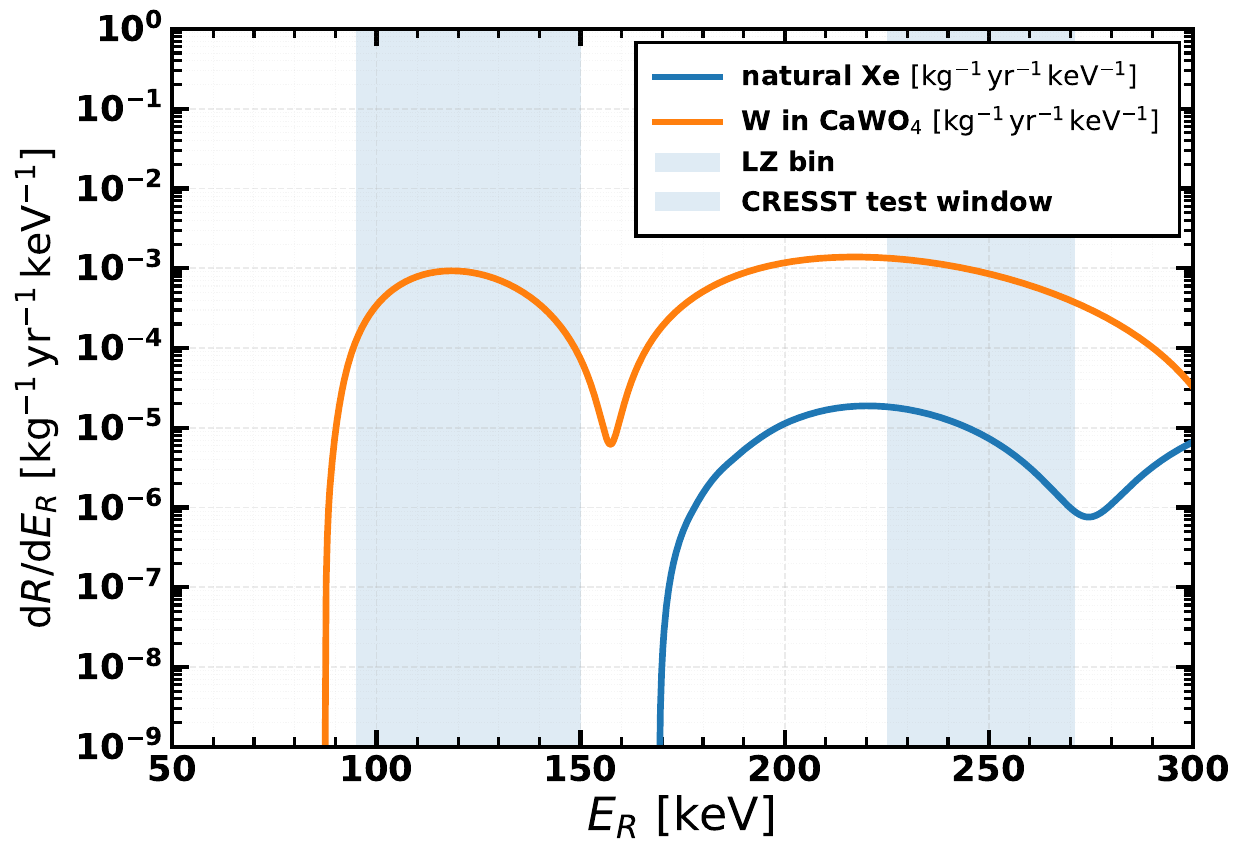}
\caption{Benchmark recoil spectra for natural xenon and for tungsten in a CaWO$_4$ target.  Tungsten opens the same endothermic transition at substantially lower recoil energy, but the fixed weak normalization keeps the absolute rate modest.}
\label{fig:xew}
\end{figure}

For tungsten we use natural isotopic abundances and the same coherent weak charge as in Eq.~(\ref{eq:sigmaA}).  Because a CRESST module is CaWO$_4$, the tungsten number density per kilogram is one W nucleus per CaWO$_4$ molecule rather than that of a pure-W target.  Figure~\ref{fig:xew} compares the recoil spectra per unit detector mass.

Following the prospective interval proposed in Ref.~\cite{Su2026}, we integrate the tungsten component over
\begin{equation}
95<E_R<150~\keV.
\end{equation}
For the thermal benchmark and unit efficiency we obtain
\begin{equation}
R_{95-150}^{\rm W}
\simeq3.15\times10^{-2}
~\mathrm{events~kg^{-1}~yr^{-1}}.
\label{eq:RW}
\end{equation}
A background-free 90\% C.L. reference of 2.3 events would therefore require
\begin{equation}
{\cal E}_{2.3}^{\rm W}\simeq72.9~\mathrm{kg~yr}
\simeq26.6~\mathrm{tonne~day},
\label{eq:EW}
\end{equation}
again before realistic nuclear-recoil acceptance and background are included.

This result is significantly more demanding than the generic benchmark in Ref.~\cite{Su2026}.  That analysis used an isoscalar nucleon cross section $\bar\sigma_n=10^{-37}\,\mathrm{cm^2}$ and found $R_{95-150}\simeq0.515$ events kg$^{-1}$ yr$^{-1}$, implying about 1.63 tonne-days for 2.3 ideal events.  The difference is physical: our electroweak-doublet model has a fixed neutron-scale cross section near $1.9\times10^{-39}\,\mathrm{cm^2}$ and the nuclear amplitude is proportional to the weak charge rather than $A$.

The planned CRESST upgrade discussed in Refs.~\cite{CRESSTUpgrade,Su2026} may reach approximately 1.5 tonne-days after three years.  At the present benchmark Eq.~(\ref{eq:RW}) corresponds to only about
\begin{equation}
N_{\rm CRESST}(1.5~\mathrm{tonne~day})\simeq0.13
\end{equation}
ideal signal events.  A null result at that exposure would therefore not yet exclude the weak geometric model.  Conversely, several events in that window would favor a stronger interaction than the minimal $Z$-exchange realization considered here.

\begin{figure}[t]
\includegraphics[width=\columnwidth]{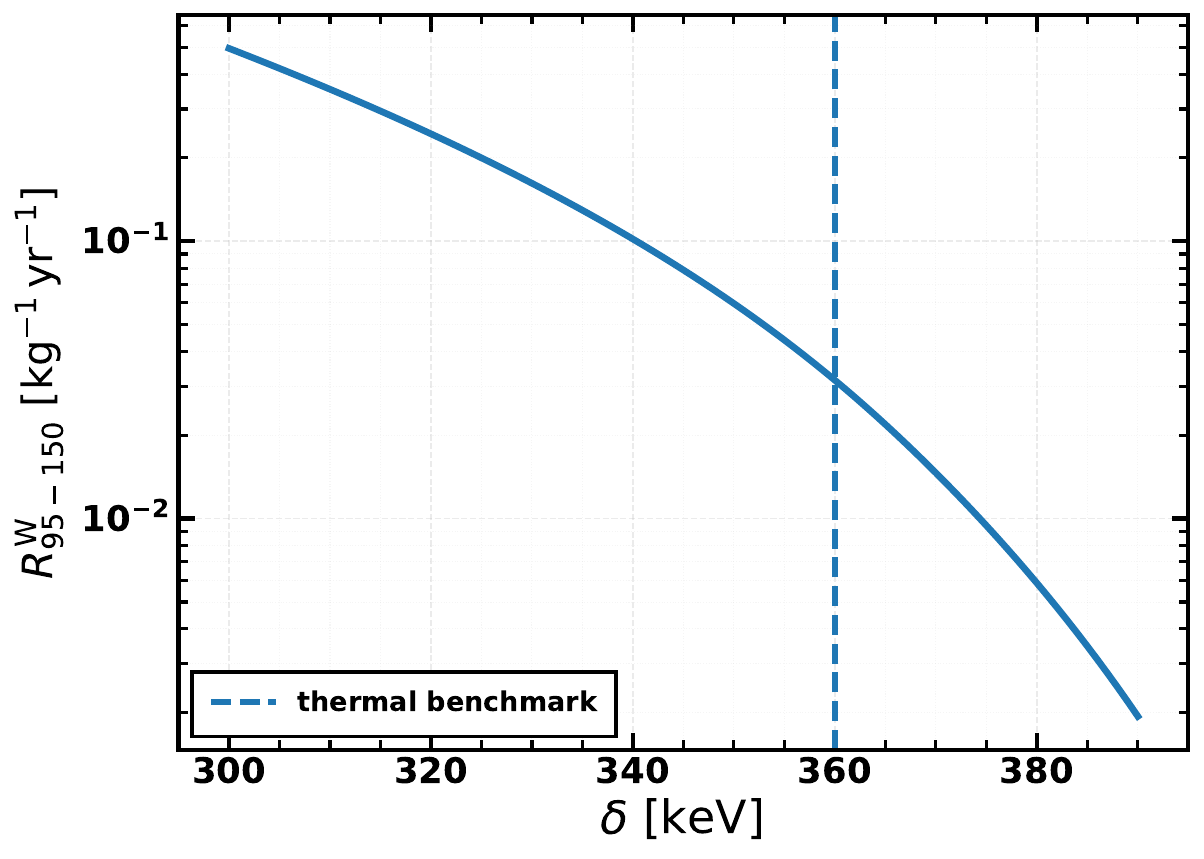}
\caption{Integrated tungsten rate in the illustrative 95--150 keV CRESST interval as a function of the inelastic splitting, for $m_\chi=1.1$ TeV and the fixed weak interaction.}
\label{fig:cresstrate}
\end{figure}

Figure~\ref{fig:cresstrate} shows the strong dependence of the tungsten rate on the inelastic splitting.  As $\delta$ approaches the kinematic ceiling, the available halo phase space shrinks rapidly and the rate in the fixed 95--150 keV interval falls correspondingly.

\section{Systematics and scope}
\label{sec:systematics}

The benchmark probes the extreme tail of the Galactic velocity
distribution. Small changes in $v_{\rm esc}$, the Earth speed, or
non-Maxwellian substructure can therefore modify the absolute rate
much more strongly than in conventional elastic WIMP searches.
Equations~(\ref{eq:velocitydist})--(\ref{eq:etaSHM}) make explicit
the particular Standard Halo Model convention adopted in our
numerical calculation. The mean inverse speed $\eta(v_{\min})$ is
therefore not a universal function, but depends on the assumed
Galactic velocity distribution.
The likelihood contours in Fig.~\ref{fig:like} should consequently
be interpreted with this astrophysical dependence in mind. A future
analysis with several high-energy events should profile the relevant
astrophysical nuisance parameters rather than fixing a single
Standard Halo Model.

At $E_R\sim250$ keV in xenon, the momentum transfer is of order a few
hundred MeV. A Helm form factor provides a transparent baseline
description, but it is not the final word at such large momentum
transfer. LZ predicts nonrelativistic EFT spectra using one-body
nuclear-response information together with a detector simulation.
The same issue is relevant for tungsten, for which the nuclear
structure at large momentum transfer is target dependent. The
kinematic conclusions summarized in Table~\ref{tab:reach} are
insensitive to the detailed form-factor prescription, whereas the
absolute rates in Eqs.~(\ref{eq:RW})--(\ref{eq:EW}) are not.

There are corresponding limitations associated with the statistical
treatment. Equation~(\ref{eq:like}) compresses the LZ information
into two recoil-energy bins and neglects background uncertainties.
It is useful for identifying phenomenologically viable regions and
for direct comparison with Ref.~\cite{Su2026}, but it should not be
confused with the official LZ collaboration profile likelihood.
Likewise, the CRESST estimate assumes unit efficiency and vanishing
background. The exposure quoted in Eq.~(\ref{eq:EW}) should therefore
be regarded as an idealized benchmark for a future detector-level
analysis rather than as an experimental sensitivity claim.

\begingroup

The Dark-Dimension interpretation introduces an additional layer of
model dependence beyond the direct-detection calculation itself.
The low-energy phenomenology depends primarily on $m_\chi$, $\delta$,
and the weak interaction, whereas the higher-dimensional embedding
also involves the bulk mass $M$, the compactification radius $R$, and
the corresponding Kaluza--Klein scale
\begin{equation}
m_{\rm KK}\equiv\frac{1}{R}.
\end{equation}
Using the same notation throughout the five-dimensional construction,
the neutral bulk excitations satisfy
\begin{equation}
m_n^2
=
M^2+\frac{n^2}{R^2}
=
M^2+n^2m_{\rm KK}^2 .
\label{eq:neutralKK}
\end{equation}
Because this tower is electroweak neutral, it avoids the immediate
charged-KK problem that would arise if the electroweak doublets
$D_{1,2}$ themselves propagated in a micron-size extra dimension.

Nevertheless, the brane--bulk couplings in
Eq.~(\ref{eq:braneYuk}) must remain sufficiently small to satisfy
electroweak-precision, invisible-width, and cosmological constraints.
A complete ultraviolet realization must also specify the boundary
conditions of the neutral bulk field, the stabilization mechanism
for the compactification radius $R$, and the microscopic origin of
the distant-boundary $\Delta D=2$ interaction. These ingredients can
modify the non-exponential prefactor contained in $\delta_{\rm UV}$.
In addition, the use of the static brane--to--brane kernel in
Eq.~(\ref{eq:propagator}) is an effective matching assumption. A
fully specified five-dimensional theory should evaluate the complete
momentum-dependent bulk propagator with the chosen boundary conditions
and verify the induced Majorana self-energy of the TeV brane state.
Within the effective description adopted here, these ultraviolet details
do not change the characteristic geometric suppression,
\begin{equation}
\delta
=
\delta_{\rm UV}e^{-2\pi MR}
=
\delta_{\rm UV}
\exp\!\left(-2\pi\frac{M}{m_{\rm KK}}\right).
\label{eq:deltaKK_scope}
\end{equation}

The construction should therefore be interpreted in two layers.
The pseudo-Dirac electroweak doublet and its direct-detection
predictions constitute the phenomenological core of the analysis,
while the Dark-Dimension sector provides a geometric interpretation
of the small Majorana splitting and relates the compactification
radius to the vacuum-energy scale. A complete treatment of the
cosmological evolution and phenomenology of the neutral KK tower
is beyond the scope of the present work and is left for future study.

\endgroup

Finally, an electroweak doublet with mass near $1.1$ TeV lies close
to the familiar thermal Higgsino mass scale. The precise relic
abundance depends on the charged--neutral spectrum, coannihilation,
and any additional states present in the ultraviolet completion.
Collider searches for the charged partner and indirect searches
therefore provide complementary information
\cite{NagataShirai2015,FanReece2026,WuZhangZhu2026}.
These constraints are complementary to the geometric question
considered here: once a pseudo-Dirac electroweak doublet is present,
the fifth-dimensional construction provides a mechanism for
generating the small splitting that controls the direct-detection
phenomenology.

\section{Conclusions}

We have studied a pseudo-Dirac electroweak-doublet dark-matter
scenario motivated by the extended high-recoil LZ analysis and
embedded it in a five-dimensional Dark-Dimension framework. In this
construction the observed vacuum-energy scale sets the characteristic
compactification radius, while dark-number violation is communicated
through a neutral bulk state. The resulting Majorana splitting is
geometrically suppressed,
\begin{equation}
\delta=\delta_{\rm UV}e^{-2\pi MR},
\end{equation}
so that the few-hundred-keV scale relevant for inelastic direct
detection can arise from an order-one value of $MR$ rather than being
introduced as an independent low-energy parameter. For
$\delta_{\rm UV}=1$ GeV and $\delta=360$ keV, one finds
$MR\simeq1.26$.

The low-energy phenomenology is particularly predictive because the
neutral weak current is dominantly off diagonal and the scattering
normalization is fixed by the electroweak interaction. Our xenon
analysis gives a shallow likelihood minimum near
$(m_\chi,\delta)\simeq(1.35~{\rm TeV},366~{\rm keV})$, while the
thermally motivated electroweak-doublet benchmark
$(1.1~{\rm TeV},360~{\rm keV})$ lies very close to this minimum.
The latter predicts approximately $(0.32,0.92)$ events in the
$160$--$200$ and $225$--$271$ keV intervals, respectively, and
therefore provides a representative point for the multi-target study.

A characteristic consequence of the endothermic threshold is the
strong dependence on the target nucleus. At the reference point,
scattering is kinematically forbidden on F, Si, Ar, Ca, and Ge, while
I, Xe, and W remain accessible. In particular, the xenon signal turns
on only near $191$ keV, naturally emphasizing the extended LZ recoil
window, whereas tungsten remains accessible from substantially lower
recoil energies. For CaWO$_4$ we find an integrated tungsten rate of
about $3.15\times10^{-2}$ events kg$^{-1}$ yr$^{-1}$ in the
$95$--$150$ keV interval, corresponding to an idealized exposure of
roughly $26.6$ tonne-days for 2.3 signal events.

The main outcome of this work is therefore a direct connection between
higher-dimensional geometry and high-recoil dark-matter phenomenology.
The compactification scale fixes the geometric setting, the bulk
propagation determines the pseudo-Dirac splitting, and the resulting
inelastic threshold predicts a distinctive hierarchy among detector
targets. Future high-energy xenon data, especially when combined with
heavy-target measurements such as tungsten, can test whether a
persistent recoil population is consistent with this geometric
electroweak-doublet interpretation.
\appendix

\section{Recoil endpoints}

Solving $v_{\min}(E_R)=v_{\rm max}$ gives the kinematic endpoints
\begin{equation}
E_R^{\pm}=
\frac{\mu_{\chi A}^2}{2m_A}
\left[
v_{\rm max}\pm
\sqrt{v_{\rm max}^2-\frac{2\delta}{\mu_{\chi A}}}
\right]^2,
\label{eq:endpoints}
\end{equation}
provided Eq.~(\ref{eq:dmax}) is satisfied.  These endpoints were used to construct Fig.~\ref{fig:windows} and Table~\ref{tab:reach}.

\end{document}